\documentclass[10pt]{llncs}

\usepackage{listings}
\usepackage{graphicx}
\usepackage{xcolor}   
\usepackage{todo}
\usepackage{hyperref}

\definecolor{dBlue}{rgb}{0.0, 0.0, 0.65}
\definecolor{dkred}{rgb}{0.65, 0.0, 0}
\definecolor{dkgreen}{rgb}{0, 0.65, 0}

\newtheorem{thm}{Theorem}[section]
\newtheorem{prop}[thm]{Proposition}

\newcommand{\raz}{RAZ}

\newcommand{\secref}[1]{Section~\ref{#1}}
\newcommand{\figref}[1]{Figure~\ref{fig:#1}}
\newcommand{\Figref}[1]{Figure~\ref{#1}}
\newcommand{\code}[1]{\lstinline[basicstyle=\footnotesize\ttfamily]{#1}}

\begin{document}

\title{
  The Random Access Zipper:
}
\subtitle{Simple, Purely-Functional Sequences}
\author{
  Kyle Headley
  \and
  Matthew A. Hammer
}
\institute{University of Colorado Boulder
  \\kyle.headley@colorado.edu, matthew.hammer@colorado.edu
  \\[2mm]
}

\authorrunning{Headley and Hammer}

\maketitle

\begin{abstract}
We introduce the Random Access Zipper (\raz{}), a simple,
purely-functional data structure for editable sequences.
The \raz{} combines the structure of a zipper with that of a tree:
like a zipper, edits at the cursor require constant time;
by leveraging tree structure, relocating the edit cursor in the
sequence requires log time.
While existing data structures provide these time bounds, none do so
with the same simplicity and brevity of code as the \raz{}. 
The simplicity of the \raz{}
provides the opportunity for more programmers to extend the structure
to their own needs,
and we provide some suggestions for how to do so.
\end{abstract}

\section{Introduction}
\label{intro}

The singly-linked list is the most common representation of sequences
for functional programmers.
This structure is considered a core primitive in every functional
language, and morever, the principles of its simple design recur
througout user-defined structures that are ``sequence-like''.
Though simple and ubiquitous, the functional list has a serious
shortcoming: users may only efficiently access and edit the
\emph{head} of the list. In particular, random accesses (or edits)
generally require linear time.

%

%
%
To overcome this problem, researchers have developed other data
structures representing (functional) sequences, most notably,
\emph{finger trees}~\cite{Hinze-Paterson:FingerTree}.
These structures perform well, allowing edits in (amortized) constant
time and moving the edit location in logarithmic time.
More recently, researchers have proposed the
\emph{{RRB}-Vector}~\cite{RRBVector}, offering a balanced tree
representation for immutable vectors.
Unfortunately, these alternatives lack the simplicity and
extensibility of the singly-linked list.

In this paper, we introduce the \emph{random access zipper}, or \raz{}
for short.
Like the common linked list, the \raz{} is a general-purpose data structure
for purely-functional sequences.
The \raz{} overcomes the performance shortcomings of linked lists by
using probabilistically-balanced trees to make random access
efficient (expected or amortized logarithmic time).
The key insight for balancing these trees comes from \cite{PughTe89},
which introduces the notion of probabilistically-chosen \emph{levels}.\footnote{
  In short, these
  levels represent the heights of uniformly randomly chosen nodes in a
  full, balanced binary tree. See \secref{tech}. }
%
%
To edit sequences in a persistent (purely-functional) setting, the \raz{} also
incorporates the design of a zipper \cite{Huet97}, which provides the
notion of a \emph{cursor} in the sequence. A cursor focuses edits on (or
near) a distinguished element.
The user may move the cursor locally (i.e., forward and backward, one
element at a time), or globally (i.e., based on an index into the
sequence), which provides random access to sequence elements.
%


The \raz{} exposes two types to the user, \code{'a tree} and \code{'a
  zip}, which respectively represent a unfocused and focused sequence
of elements (of type~\code{'a}).
The \raz{} exposes the following interface to the user based on these
types:

\begin{center}
\begin{tabular}{|l@{~:~}l@{~~~~}l|}
\hline
\textbf{Function}  & \textbf{Type} & \textbf{Time Complexity}
\\
\hline
\texttt{focus}   & \texttt{'a tree -> int -> 'a zip} & $O(\log n)$~expected
\\
\texttt{unfocus} & \texttt{'a zip -> 'a tree}        & $O(\log n + m\cdot\log^2m)$~expected
\\
\hline
\texttt{insert}  & \texttt{dir -> 'a -> 'a zedit}    & $O(1)$~worst-case
\\
\texttt{remove}  & \texttt{dir -> 'a zedit}          & $O(1)$~amortized
\\
\texttt{replace} & \texttt{'a -> 'a zedit}           & $O(1)$~worst-case
\\
\texttt{move}    & \texttt{dir -> 'a zedit}          & $O(1)$~amortized
\\
\texttt{view}    & \texttt{'a zip -> 'a}             & $O(1)$~worst-case
\\
\hline
\end{tabular}
\end{center}

The second and third columns of the table respectively report the type
(in OCaml) and time complexity for each operation; we explain each in
turn.

Function~\code{focus} transforms a tree into a zipper, given a
position in the sequence on which to place the zipper's cursor.
It runs in expected logarithmic time, where $n$ is the number of
elements in the sequence.  We use expected analysis for this function
(and the next) since the tree is balanced probabilistically.

Function~\code{unfocus} transforms a (focused) zipper back to an
(unfocused) tree; its time complexity $O(\log n +
m\cdot\log^2m)$~depends on the length of the sequence~$n$, as well as
$m$, the number of zipper-based edits since the last refocusing.
We summarize those possible edits below.
Assuming that the number~$m$ is a small constant, the complexity of
\code{unfocus}~is merely $O(\log n)$; when $m$ grows to become
significant, however, the current design of the \raz{} performs more
poorly, requiring an additional expected $O(\log^2m)$ time to process
each edit in building the balanced tree.
To overcome this problem, the user can choose to refocus more often
(e.g., after each edit, if desired).

Functions~\code{insert}, \code{replace}, \code{remove}, \code{move}
each transform the zipper structure of the \raz{}.
We abbreviate their types using type~\code{'a zedit}, which we define
as the function type~\code{'a zip -> 'a zip}.
Function~\code{insert} inserts a given element to the left or right of
the cursor, specified by a direction of type~\code{dir}.
Function \code{remove} removes the element in the given direction. Its
time complexity is amortized, since removal may involve decomposing
subtrees of logarithmic depth; overall, these
costs are amortized across edits that require them.
Function~\code{replace} replaces the element at the cursor with the
given element.
Function~\code{move} moves the cursor one unit in the specified
direction; just as with \code{remove}, this operation uses amortized
analysis.
Finally, function~\code{view} retrieves the element currently focused
at the cursor (in constant time).

In the next section, we give an in-depth example of using the
operations described above.
In particular, we illustrate how the \raz{} represents the sequence as
a tree and as a zipper, showing how the operations construct and
transform these two structures.

In \secref{tech}, we present our implementation of the \raz{}.  It requires well under 200 lines of OCaml, which is publicly available:
\begin{center}
\url{https://github.com/cuplv/raz.ocaml}
\end{center}
This code includes ten main functions that work over a simple set of
datatypes for trees, lists of trees, and zippers, as defined in
\secref{tech}.
In contrast, the finger tree \cite{Hinze-Paterson:FingerTree} requires
approximately 800 lines in the OCaml ``Batteries Included'' repo
\cite{batteries-repo} to provide similar functionality.
%
%

We evaluate the \raz{} empirically in \secref{eval}.
In particular, we report the time required to build large sequences
by inserting random elements into the sequence at random positions.
Our evaluation demonstrates that the \raz{} is very competitive with
the finger tree implementation mentioned above, despite the simplicity
of the \raz{} compared with that of the finger tree.

In \secref{discuss}, we discuss the design decisions we considered for
this implementation and exposition of the~\raz{}.
We also discuss future enhancements that build on this design.

\secref{related} discusses related work, and in particular,
alternative structures for persistent sequences.
We give a deeper comparison to finger trees, and explain why other
balanced trees designed for sets~(e.g., red-black trees, or splay
trees, etc) are inappropriate for representing sequences.

\section{Example}
\label{sec:overview}
\label{sec:example}
\label{example}

In this section, we give a detailed example of using the \raz{}
interface introduced above, and illustrate the internal tree and list
structures of the \raz{} informally, using pictures.
In the next section, we make the code for these functions precise.

\begin{figure}
\begin{center}
Elements and levels for our example sequence:
\\
\includegraphics[width=3.5in]{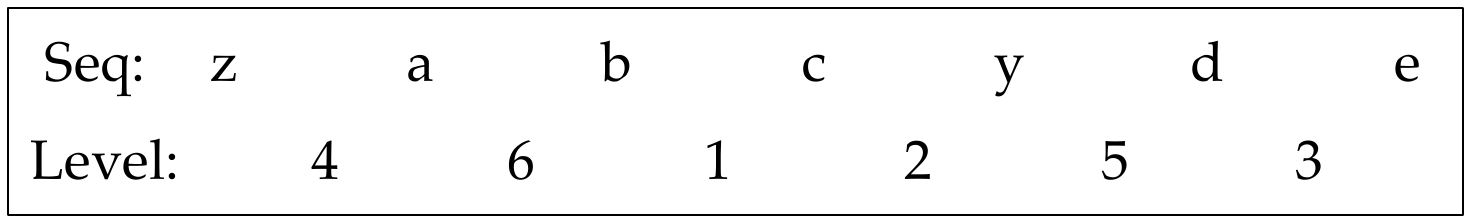}
\\
As a tree, levels are internal and determine heights:
\\
\includegraphics[width=3.5in]{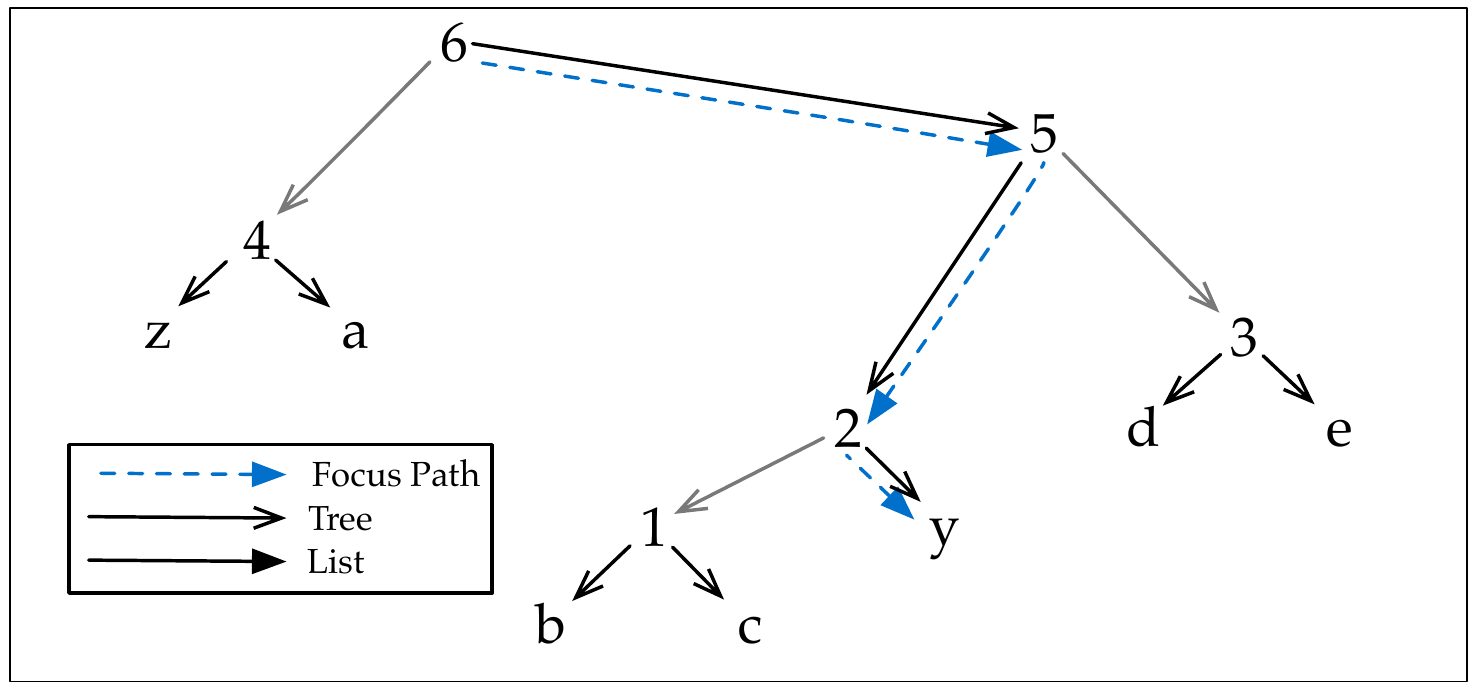}
\\
Zipper focused on $y$. Three (unfocused) sub-trees remain:
\\
\includegraphics[width=3.5in]{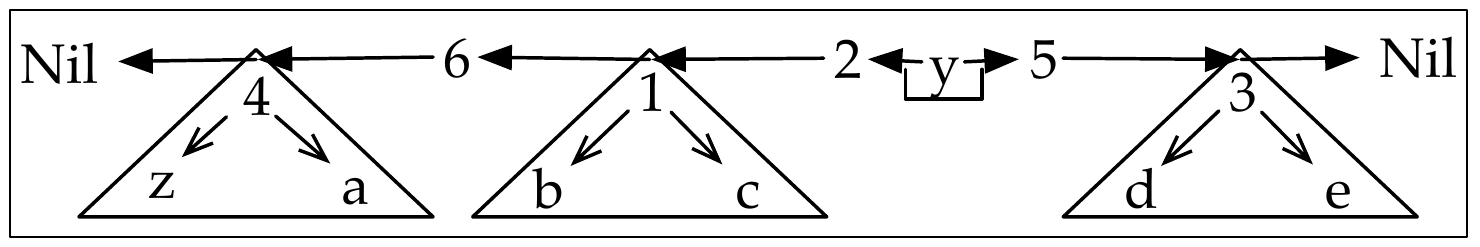}
\end{center}
\caption{The \raz{} represents the sequence of elements
  $\left<z,a,b,c,y,d,e\right>$ interposed with levels 1-6 (first
  image); these levels uniquely determine a balanced tree (second
  image) that permits log-time focusing on element~$y$ (third image).}
\label{focused-raz-from-sequence}
\begin{center}
\includegraphics[width=3.5in]{overviewtrees2-focused}
\\
Remove the level $2$, to the left of the cursor:
\\
\includegraphics[width=3.5in]{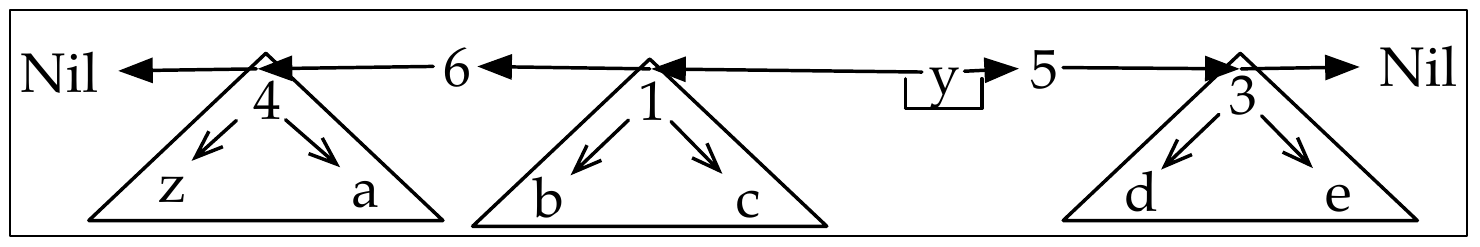}
\\
Trim the left of the cursor, deconstructing its rightmost path
\\
\includegraphics[width=3.5in]{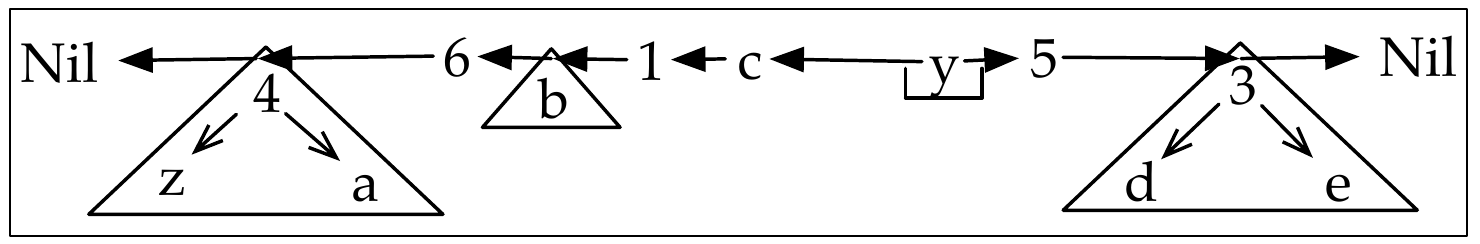}
\\
Remove the $c$
\\
\includegraphics[width=3.5in]{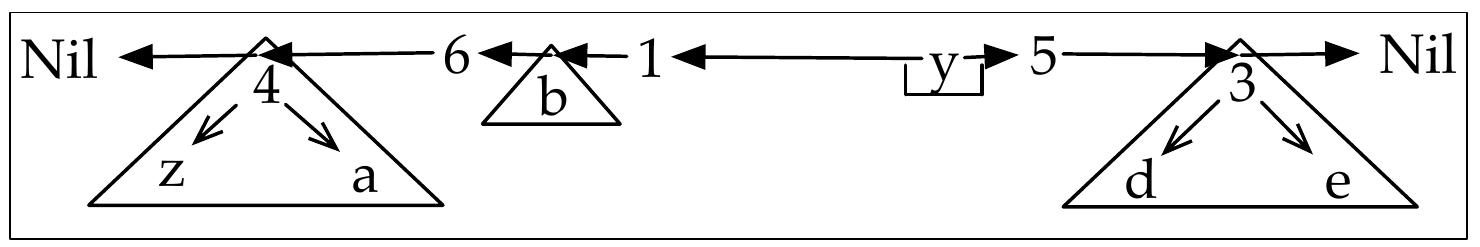}
\end{center}
\caption{An example of editing a focused \raz{}: Remove the $c$ to the left of the cursor by removing level $2$ (second image), trimming the left tree (third image), and removing the $c$ (last image)}
\label{raz-remove-left}
\end{figure}

\begin{figure}
\begin{center}
\includegraphics[width=3.5in]{overviewtrees2-removedc}
\\
Store element $y$ in the sequence to be unfocused:
\\
\includegraphics[width=3.5in]{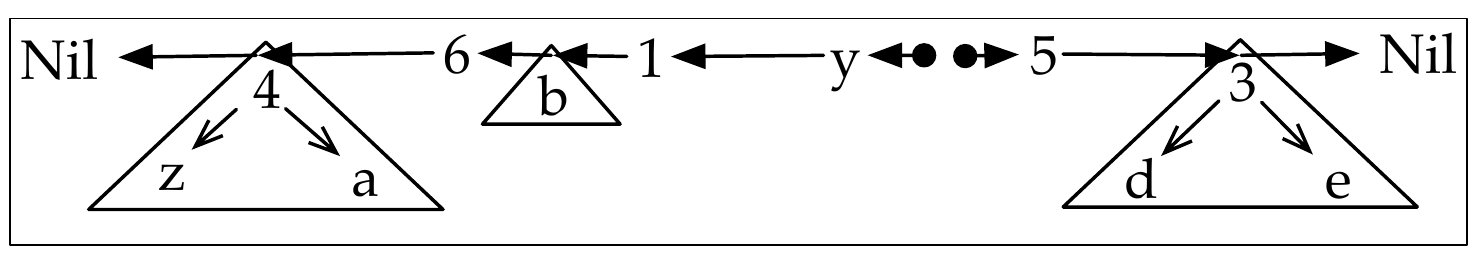}
\\
Append the left and right sequences, represented as trees:
\\
\includegraphics[width=3.5in]{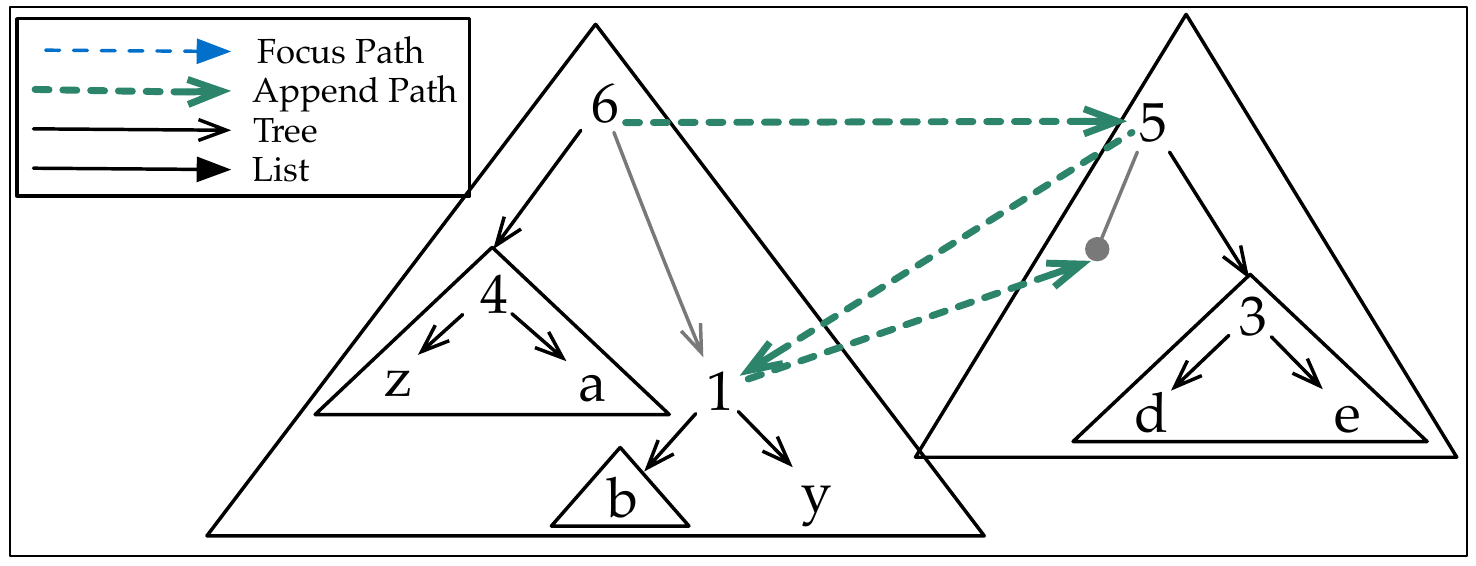}
\\
Result of appending:
\\
\includegraphics[width=3.5in]{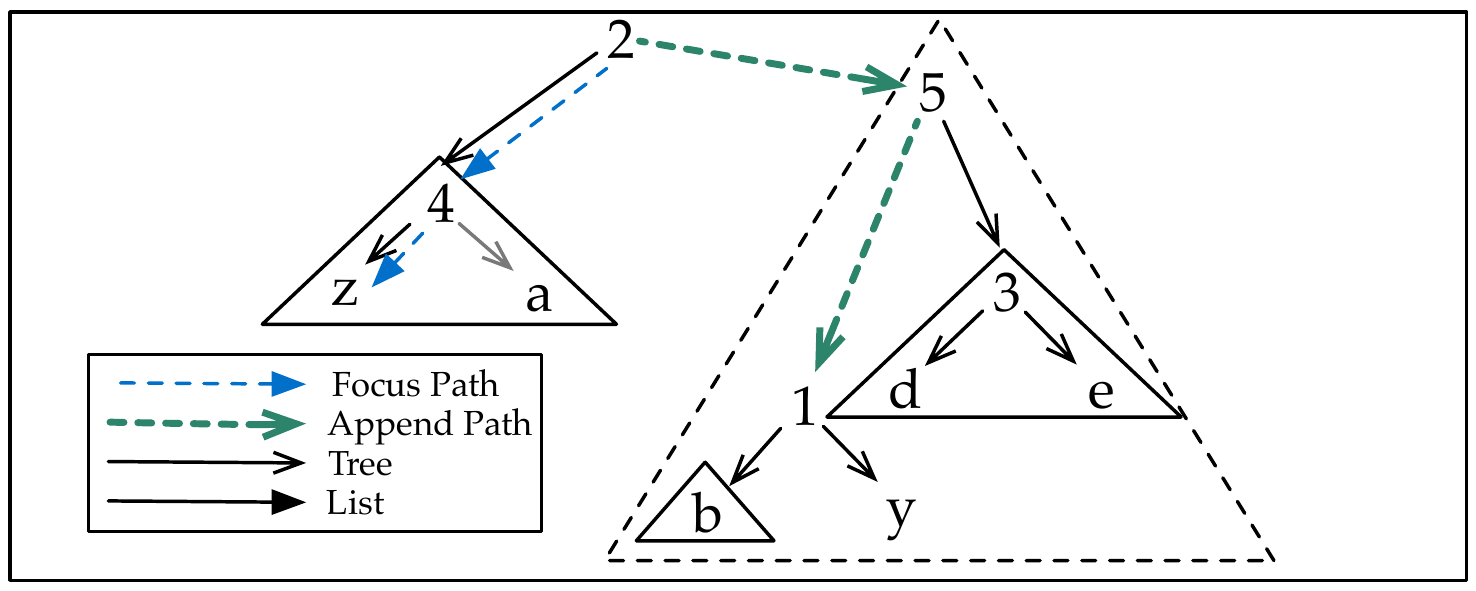}
\end{center}
\vspace{-10px}
\caption{Unfocus the \raz{}}
\label{raz-append}
\begin{center}
\includegraphics[width=3.5in]{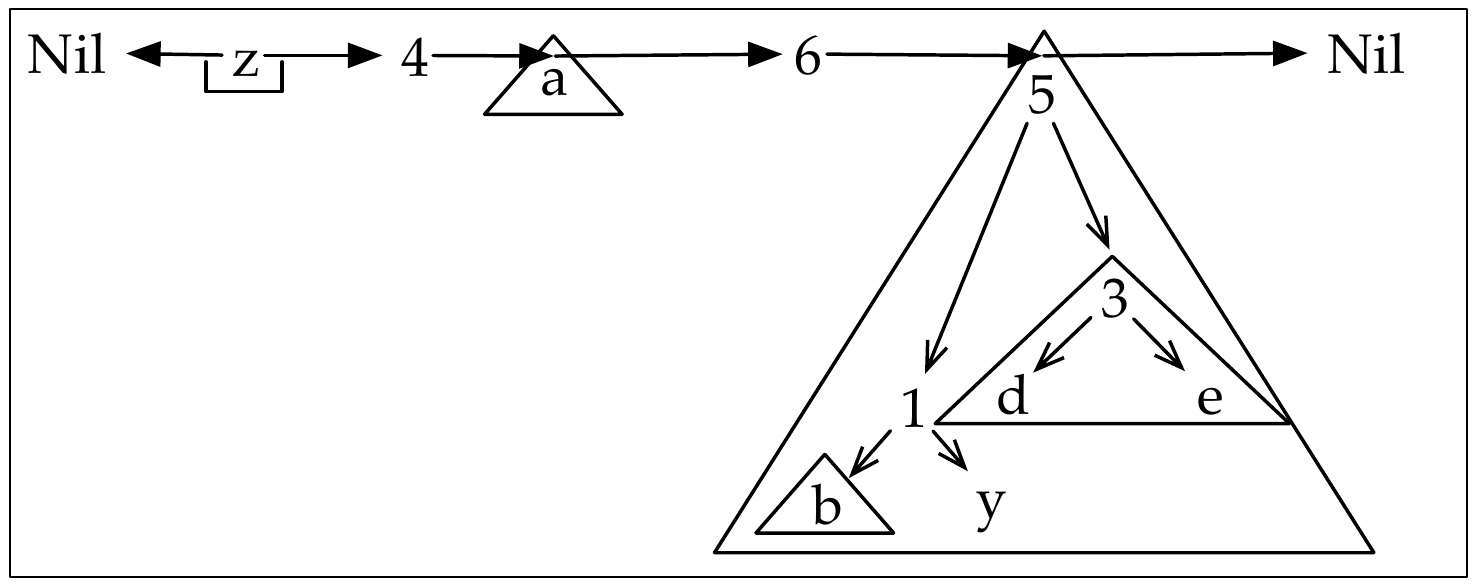}
\end{center}
\caption{Focus on the first element, $z$, creating left and right lists of (unfocused) subtrees.}
\label{raz-refocus}
\end{figure}

Consider the sequence of seven elements~$\left<z,a,b,c,y,d,e\right>$.
In the example that follows, the programmer uses the \raz{} to perform
four operations over this sequence, editing it (functionally) in the
process:
\begin{itemize}
\item She uses \code{focus} to place the cursor at offset~4 (element~$y$), 
\item she uses \code{remove} to remove the element to the left of the cursor (element~$c$), 
\item she uses \code{unfocus} in anticipation of refocusing, and
\item she uses \code{focus} to place the cursor at the sequence's start (element~$z$).
\end{itemize}

\Figref{focused-raz-from-sequence} (first image, top) shows the
sequence of elements represented by a \raz{}.
As explained further in \secref{tech}, the \raz{} interposes
randomly-chosen \emph{levels} as meta-data in the sequence; we show
these levels,~$\left<4,6,1,2,5,3\right>$, interposed below the
sequence elements.
%
%
%
%
The second image shows the tree form of the \raz{}, whose structure is
determined by the sequence elements and levels above.
This tree represents an unfocused \raz{}.

Next, the programmer uses \code{focus} to place the cursor into the
tree, in order to edit its elements.
The third image in \Figref{focused-raz-from-sequence} shows the zipper
that results from focusing the sequence on element~$y$.
As can be seen, this structure consists of left and right lists that
each contain levels and unfocused subtrees from the original balanced
tree.
The focusing algorithm produces this zipper by descending the balanced
tree along the indicated focus path (second image of
\Figref{focused-raz-from-sequence}), adding names and subtrees along
this path to the left and right lists.
Notice that the elements nearest to the cursor consist of the subtrees
at the end of this path; in expectation, these lists order subtrees in
ascending size.

After focusing on element~$y$, the programmer uses the \code{remove}
function.
\Figref{raz-remove-left} shows the three steps for removing the $c$ to
the left of cursor.
First, we remove the level $2$ from the left of the cursor, making $c$
the next element to the immediate left of the cursor (the second
image).
Next, since the $c$ resides as a leaf in an unfocused tree, we
\emph{trim} this left tree by deconstructing its rightmost path (viz.,
the path to $c$).
\secref{tech} explains the trim operation in detail.
Finally, with $c$ exposed in the left list, we remove it (fourth
image).

After removing element~$c$, the programmer unfocuses the \raz{}.
Beginning with the final state of \Figref{raz-remove-left},
\Figref{raz-append} illustrates the process of unfocusing the
sequence.
First, we add element $y$ to the left list, storing its position in
the sequence (second image).
Next, we build trees from the left and right lists as follows: For
each list, we fold its elements and trees, appending them into
balanced trees; as with the initial tree, we use the levels to
determine the height of internal nodes (third image).
Having created two balanced trees from the left and right lists, we
append them along their rightmost and leftmost paths, respectively;
again, the append path compares the levels to determine the final
appended tree (fourth image).

Finally, the programmer refocuses the \raz{}.
In \Figref{raz-refocus} (as in \Figref{focused-raz-from-sequence}), we
descend the focus path to the desired element, this time $z$. As
before, this path induces left and right lists that consist of levels
and unfocused subtrees.

\section{Technical Design}
\label{tech}

The full implementation of the \raz{} in OCaml consists of about 170 lines. 
%
%
In this section, we tour much of this code, with type signatures for
what we elide for space considerations.

\begin{figure}
\begin{lstlisting}[basicstyle=\footnotesize\ttfamily]
type lev = int   (* tree level *)
type dir = L | R (* directions for moving/editing *)

type 'a tree = (* binary tree of elements *)
              | Nil 
              | Leaf of 'a
              | Bin  of lev * int * 'a tree * 'a tree

type 'a tlist = (* list of elements, levels and trees *)
              | Nil 
              | Cons  of 'a      * 'a tlist
              | Level of lev     * 'a tlist
              | Tree  of 'a tree * 'a tlist
type 'a zip  = ('a tlist * 'a * 'a tlist) (* tlist zipper *)
\end{lstlisting}
\caption{\raz{} defined as a zipper of tree-lists.}
\label{fig:types}
\end{figure}

\figref{types} lists the type definitions for the \raz{} structure, which
is stratified into three types: \code{tree}, \code{tlist}, and \code{zip}.
The \code{tree} type consists of (unfocused) binary trees, where
leaves hold data, and where internal binary nodes hold a level~\code{lev}
and total element count of their subtrees (an \code{int}).
The \code{tlist} type consists of ordinary list structure, plus two
\code{Cons}-like constructors that hold \code{lev}s and \code{tree}s
instead of ordinary data.
Finally, a (focused) \code{zip} consists of a left and right
\code{tlist}, and a focused element that lies between them.

\paragraph{Levels for probabilistically-balanced trees.}
As demonstrated in the code below for \code{append}, the levels
associated with each \code{Bin} node are critical to maintaining
balanced trees, in expectation.
This property of balance is critical to the time complexity
bounds given for many of the \raz{}'s operations, including focusing,
unfocusing and many local edits.

The key insight is choosing these levels from a \emph{negative
  binomial distribution}; intuitively, drawing random numbers from
this distribution yields smaller numbers much more often (in
expectation) than larger numbers.
More precisely, drawing the level~1 is twice as likely as drawing the
level~2, which is twice as likely as level~3, and so on.
This means that, in expectation, a sequence of levels drawn from this
distribution describes the sizes of subtrees in a
perfectly-balanced binary tree.
As described in \secref{related}, this insight comes from
\cite{PughTe89}, who define the notion of level in a related context.

Fortunately, we can choose these levels very quickly, given a source
of (uniformly) random numbers and a hash function.
We do so by hashing a randomly-chosen number, and by counting the
number of consecutive zeros in this hash value's least-significant
bits.

\begin{figure}[h!]
\begin{lstlisting}[basicstyle=\footnotesize\ttfamily]
let focus : 'a tree -> int -> 'a zip =
fun t p ->
  let c = elm_count t in
  if p >= c || p < 0 then failwith "out of bounds" else
  let rec loop = fun t p (l,r) -> match t with
   | Nil -> failwith "internal Nil"
   | Leaf(elm) -> assert (p == 0); (l,elm,r)
   | Bin(lv, _, bl, br) -> 
     let c = elm_count bl in
     if p < c then loop bl p       (l,Level(lv,Tree(br,r)))
     else          loop br (p - c) (Level(lv,Tree(bl,l)),r)
  in loop t p (Nil,Nil)
\end{lstlisting}
\caption{The \code{focus} operation transforms a \code{tree} into a \code{zip}.}
\label{fig:focus}
\end{figure}

\paragraph{Focusing the \raz{}.}
The \code{focus} operation in \figref{focus} transforms an unfocused
tree to a focused zipper.
Given an index in the sequence, \texttt{p}, and
an $O(1)$-time \code{elm_count} operation on sub-trees, the
inner~\code{loop} recursively walks through one path of
\code{Bin}~nodes until it finds the desired \code{Leaf} element.
At each recursive step of this walk, the~\code{loop} accumulates un-walked subtrees in the pair~\code{(l,r)}.
In the base case, \code{focus} returns this accumulated \code{(l,r)}~pair as a \code{zip} containing the located leaf element.
\begin{prop}
Given a tree~\code{t} of depth~$d$, and an $O(1)$-time implementation of
\code{elm_count}, the operation~\code{focus t p} runs in $O(d)$
time.
\end{prop}

\begin{figure}[h!]
\begin{lstlisting}[basicstyle=\footnotesize\ttfamily]
let head_as_tree : 'a tlist -> 'a tree
let tail : 'a tlist -> 'a tlist

let grow : dir -> 'a tlist -> 'a tree =
 fun d t ->
  let rec loop = fun h1 t1 ->
    match t1 with Nil -> h1 | _ -> 
    let h2 = head_as_tree t1 in
    match d with
    | L -> loop (append h2 h1) (tail t1)
    | R -> loop (append h1 h2) (tail t1)  
  in grow (head_as_tree t) (tail t)

let unfocus : 'a zip -> 'a tree =
  fun (l,e,r) -> append (grow L l) (append (Leaf(e)) (grow R r))
\end{lstlisting}
\caption{Unfocusing the \raz{} using \code{append} and \code{grow}.}
\label{fig:unfocus}
\end{figure}

\paragraph{Unfocusing the \raz{}.}
\figref{unfocus} lists the \code{unfocus} operation, which transforms a
focused~\code{zipper} into an unfocused~\code{tree}.
To do so, \code{unfocus} uses auxiliary operations \code{grow} and \code{append} to construct and append trees for
the~\code{left} and \code{right} \code{tlist}~sequences that comprise the zipper.
%
%
The final steps of~\code{unfocus} consists of appending the left tree,
focused element~\code{e} (as a singleton tree), and the right tree.
In sum, the \code{unfocus} operation consists of calls to
auxiliary operations \code{grow} and \code{append}.
We explain~\code{append} in detail further below.

The \code{grow} operation uses \code{append}, and the simpler helper
function~\texttt{head\_as\_tree}, which transforms the head constructor of
a~\code{tlist} into a~\code{tree}; conceptually, it extracts the next
tree, leaf element or binary node level as \code{tree}~structure.
It also uses the function~\code{tail}, which is standard.
The \code{grow} operation loops over
successive trees, each extracted by \code{head_as_tree}, and it
combines these trees via~\code{append}.
The direction parameter~\code{d} determines whether the accumulated tree grows
from left-to-right (\code{L} case), or right-to-left (\code{R} case).
When the~\code{tlist} is~\code{Nil}, the~\code{loop}
within~\code{grow} completes, and yeilds the accumulated tree~\code{h1}.

Under the conditions stated below, \code{unfocus} is efficient,
running in polylogarithmic time for balanced trees with logarithmic
depth:

\begin{prop}
  Given a tree~\code{t} of depth~$d$, performing \code{unfocus (focus t p)} requires $O(d)$ time.
\end{prop}

We sketch the reasoning for this claim as follows.
As stated above, the operation~\code{focus t p} runs in~$O(d)$ time;
we further observe that \code{focus} produces a zipper with left and
right lists of length $O(d)$.
Likewise, \code{head_as_tree} also runs in constant time.
Next, the \code{unfocus}~operation uses \code{grow} to produce left and
right trees in $O(d)$ time. 
In general, \code{grow} makes $d$ calls to 
\code{append}, combining trees of height approaching $d$, requiring
$O(d^2)$ time. However, since these trees were placed \emph{in order} by
\code{focus}, each \code{append} here only takes constant time.
Finally, it \code{append}s these trees in $O(d)$ time.
None of these steps dominate asymptotically, so the composed operations
run in $O(d)$ time.

\begin{figure}[h!]
\begin{lstlisting}[basicstyle=\footnotesize\ttfamily]
let rec append : 'a tree -> 'a tree -> 'a tree =
  fun t1 t2 -> 
  let tot = (elm_count t1) + (elm_count t2) in
  match (t1, t2) with
  | Nil, _ -> t2 | _, Nil -> t1
  | Leaf(_), Leaf(_)       -> failwith "leaf-leaf should not arise"
  | Leaf(_), Bin(lv,_,l,r) -> Bin(lv, tot, append t1 l, r)
  | Bin(lv,_,l,r), Leaf(_) -> Bin(lv, tot, l, append r t2)
  | Bin(lv1,_,t1l,t1r), Bin(lv2,_,t2l,t2r) ->
           if lv1 >= lv2 then Bin(lv1, tot, t1l, append t1r t2)
                         else Bin(lv2, tot, append t1 t2l, t2r)
\end{lstlisting}
\caption{Append the sequences of two trees into a single sequence, as a balanced tree.}
\label{fig:append}
\end{figure}

\paragraph{Appending trees.}
The \code{append} operation in \figref{append} produces a tree whose
elements consist of the elements (and levels) of the two input trees, in order.
That is, an in-order traversal of the tree result of~\code{append t1
  t2} first visits the elements (and levels) of tree~\code{t1}, followed by
the elements (and levels) of tree~\code{t2}.
The algorithm works by traversing a path in each of its two tree
arguments, and producing an appended tree with the aforementioned
in-order traversal property.
In the last \code{Bin}~node case, the computation chooses between
descending into the sub-structure of argument~\code{t1} or
argument~\code{t2} by comparing their levels and by choosing
the tree named with the higher level.
As depicted in the example in~\Figref{raz-refocus}~(from
\secref{example}), this choice preserves the property
that~\code{Bin} nodes with higher levels remain higher in the
resulting tree.
Below, we discuss further properties of this algorithm, and
compare it to prior work.

\begin{figure}[h!]
\begin{lstlisting}[basicstyle=\footnotesize\ttfamily]
let trim : dir -> 'a tlist -> 'a tlist =
  fun d tl -> match tl with
  | Nil | Cons _ | Level _ -> tl
  | Tree(t, rest) ->
  let rec trim = fun h1 t1 -> 
    match h1 with
    | Nil -> failwith "malformed tree"
    | Leaf(elm) -> Cons(elm,t1)
    | Bin(lv,_,l,r) -> 
      match d with
      | L -> trim r (Level(lv,Tree(l,t1)))
      | R -> trim l (Level(lv,Tree(r,t1)))
  in trim t rest
\end{lstlisting}
\caption{Function~\code{trim} exposes the next sequence element.}
\label{fig:trim}
\end{figure}

\paragraph{Trimming a tree into a list.}
The \code{trim} operation in \figref{trim} prepares a \code{tlist} for
edits in the given direction \code{dir}.
It returns the same, unchanged \code{tlist} if it does not contain a
tree at its head.
If the \code{tlist} does contain a tree at its head, \code{trim}
deconstructs it recursively.
Each recursive call eliminates a \code{Bin} node, pushing the branches into the \code{tlist}.
The recursion ends when \code{trim} reaches a \code{Leaf} and pushes
it into the \code{tlist} as a \code{Cons}.

The \code{trim} operation works most efficiently when it immediately
follows a re-focusing, since in this case, the cursor is surrounded by
leaves or small subtrees, which each trim in constant time.
If the cursor moves far through the zipper, however, it can encounter
a node from high in the original tree, containing a significant
proportion of the total elements of the sequence.

These facts suggest the following propositions:

\begin{prop}
Given a direction $d$, a position $p$, a tree $t$ of size $n$, and a
\code{tlist} $l$ from one side of a zipper created by \code{focus t p}, 
 \code{trim} $d$ $l$ runs in $O(1)$ time.
\end{prop}

\begin{prop}
Given a direction $d$, a position $p$, a tree $t$ of size $n$, and a
\code{tlist} $l$ from one side of a zipper created by \code{focus t
  p}, a sequence of $k$ calls to \code{trim} $d$ $l$ composed with
\code{move d} runs in $O(k~\log~n)$ time.
\end{prop}

\begin{figure}[h!]
\begin{lstlisting}[basicstyle=\footnotesize\ttfamily]
let insert d ne (l,e,r) = match d with
  | L -> (Level(rnd_level(),Cons(ne,l)),e,r)
  | R -> (l,e,Level(rnd_level(),Cons(ne,r)))

let remove : dir -> 'a zip -> 'a zip =
  let rec remove' d s = match s with
    | Nil            -> failwith "no elements"
    | Cons(_,rest)   -> rest
    | Level(lv,rest) -> remove' d rest
    | Tree _         -> remove' d (trim d s)
  in fun d (l,e,r) -> match d with
  | L -> (remove' L l,e,r)
  | R -> (l,e,remove' R r)

let view    : dir -> 'a zip -> 'a           = ..
let replace : dir -> 'a -> 'a zip -> 'a zip = ..
let move    : dir -> 'a zip -> 'a zip       = ..

let view_cursor    : 'a zip -> 'a           = ..
let replace_cursor : 'a -> 'a zip -> 'a zip = ..
\end{lstlisting}
\caption{Zipper edits: Element insertion, element removal and several other variants.}
\label{fig:edits}
\end{figure}

\figref{edits} lists the code for inserting and removing elements from the zipper.
The function~\code{insert} uses~\code{rnd_level} to generate a random
level to accompany the newly-inserted element~\code{ne}.
Based on the removal direction, the function~\code{remove} uses an
internal helper~\code{remove'} to remove the next sequence element in
the given direction, possibly by looping.  In particular, the
\code{Cons} case is the base case that removes the element of the
\code{Cons} cell; the \code{Nil} case is erroneous, since it means that
there is no element to remove.  The two remaining cases recur
internally; specifically, the \code{Tree} case uses \code{trim},
explained above.

\figref{edits} lists the type signatures of several other zipper-based
editing functions: \code{view} accesses the next element to the right
or left of the cursor, \code{replace} replaces this element with
another given one, and \code{move} moves the cursor to focus on an
adjacent element.
Finally, \code{view_cursor} and \code{replace_cursor} are similar to
\code{view} and \code{replace}, respectively, except that they act on
the element at the cursor, rather than an element that is adjacent to
it.

\section{Evaluation}
\label{eval}

In this section we evaluate the \raz{} in comparison to a data structure with similar theoretic behavior, the finger tree \cite{Hinze-Paterson:FingerTree}, which we elaborate on in related work, \secref{related}. We demonstrate that the \raz{} is comparable in performance with this more complex structure.

\begin{figure}
\begin{center}
\includegraphics[width=0.8\columnwidth]{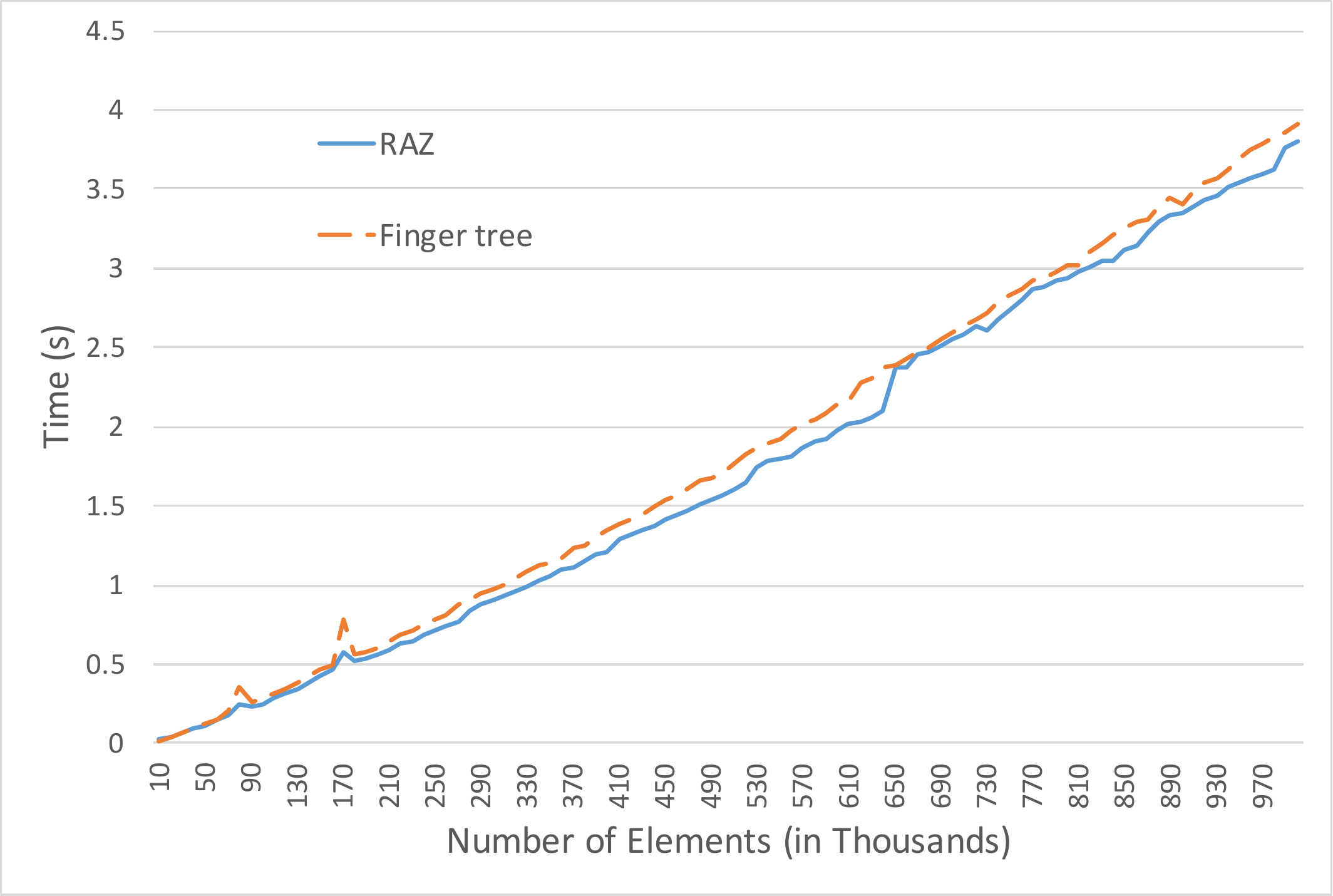}
\caption{Constructing sequences of different lengths from scratch}
\label{plot-construct}
\end{center}
\end{figure}

\paragraph{Experimental Setup.}
We present two experiments, each performed on both a \raz{} and a finger tree. In the first experiment, we construct a sequence from scratch by inserting elements, each in a randomly chosen position. Insertion into the \raz{} is by focusing, inserting a value, then unfocusing; insertion into the finger tree is by splitting, pushing a value, then appending. Upon reaching a target length, we record the total time taken. We use target lengths of 10k to 1M elements, and repeat the process for a total of five data points for each target. We plot the median of the results, shown in \Figref{plot-construct}.

\begin{figure}
\begin{center}
\includegraphics[width=0.8\columnwidth]{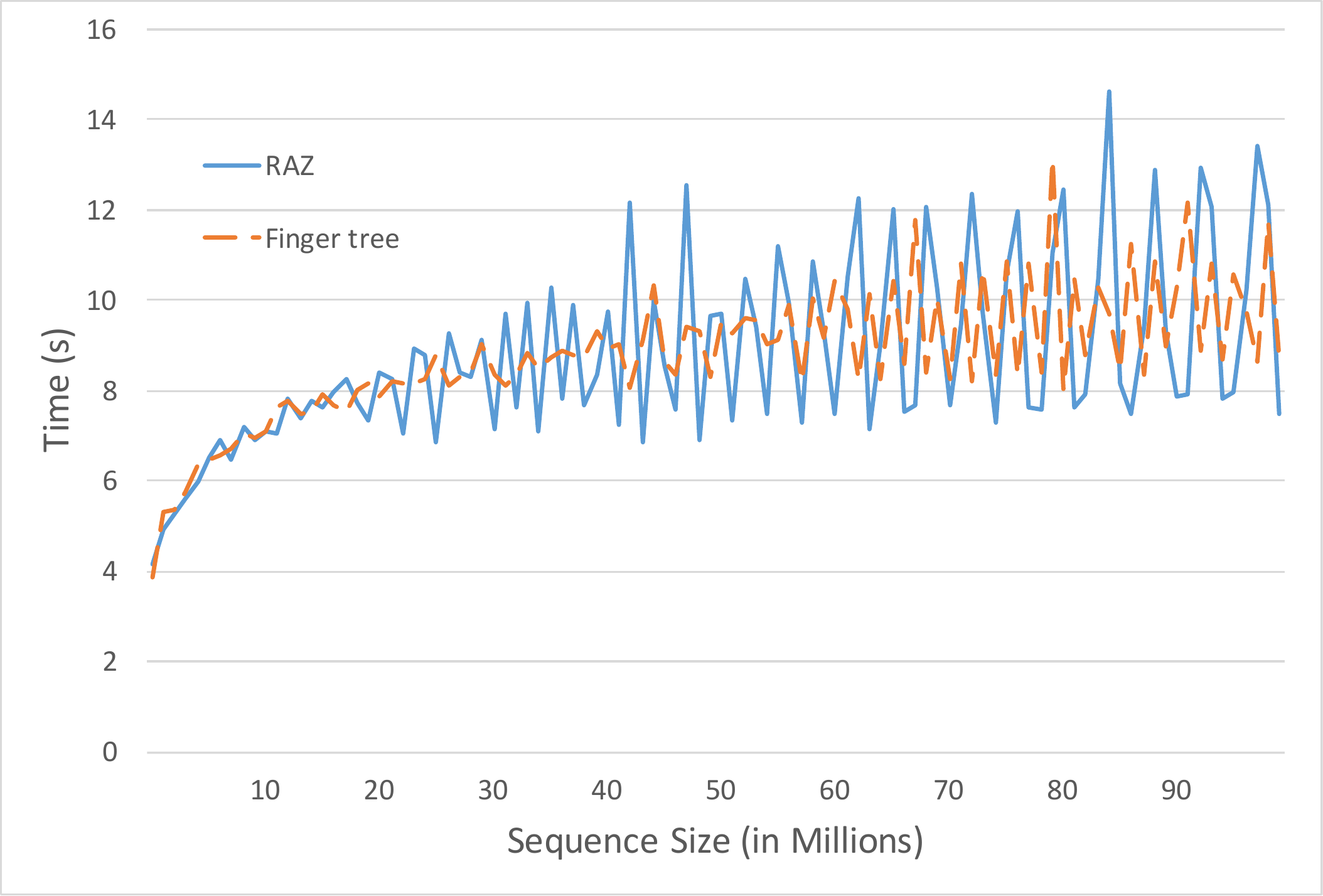}
\caption{Time taken to insert 1M elements into a sequence of varying size.}
\label{plot-insert}
\end{center}
\end{figure}

For the second experiment, we also insert elements into random positions in a sequence, but maintain a single sequence throughout the experiment. We measure the time taken for the first sequential group of 1M insertions, then the next group of 1M insertions, repeating until the sequence reaches 100M elements. We plot these measured times, shown in \Figref{plot-insert}.

We compiled our code through opam with ocamlc verion 4.02 native mode, and ran it on a 2.8 GHz Thinkpad with 16 GB of RAM running Ubuntu 16.04. We use the ``Batteries Included'' \cite{batteries-repo} code for finger trees. The results in \Figref{plot-construct} were collected by using a separate process for each measurement, and those in \Figref{plot-insert} used one process per data structure. Ocaml's \texttt{min\_heap\_size} parameter was set to 800MB. We report 100 values per data structure per plot.

\paragraph{Results.}

\Figref{plot-construct} shows the \raz{} as slightly faster than a finger tree, but maintaining the same asymptotic behavior. We average the ratio of times, showing that the \raz{} uses 5-6\% less time on average. At 500k elements the \raz{} takes 1.57s vs the finger tree's 1.71s, and at 950k elements the times are 3.54s and 3.70s, respectively. \Figref{plot-insert} shows a great variance in times. Even with a million elements added for each measurement, the plot is not smooth. This is true for the \raz{} with its probabilistic structure, but also for the more consistently structured finger trees. The average time in the last 20 entries plotted is 10.01s for the RAZ and 9.77s for the finger tree.
We suspect that garbage collection effects may be (partly) responsible for this variance, 
but determining this with certainty is beyond the scope of this initial evaluation.


\section{Discussion}
\label{discuss}

We made a number of design decisions for the \raz{}, mainly opting for simplicity of code. One decision was to use a probabilistically balanced tree rather than the strict structure of many other tree types. Storing levels eliminates the need to rebalance a tree, which eliminates some of the most complex code required. The \raz{} may not be well balanced in any local area, but it will be balanced globally, which is important for performance at scale.

An additional benefit of providing a tree with specific heights is that of a stable form. The structure of the \raz{} as a tree does not depend on the order of operations, but on the stored levels. This results in minimal changes, specifically, only the path from the adjusted elements to the root will be different. This is of great benefit for incremental computation, which is most effective for small changes. Early incremental structures by \cite{PughTe89} used elements to determine heights, and had trouble with identical values. The \raz{} sidesteps this issue with explicit stored levels for tree heights.

Another design decision we made was to provide a current element at the cursor position, rather than leaving the cursor between elements. Doing this provided about a 25\% reduction in code, removing some asymmetric logic dealing with alteration of element and level. By singling out a current item, we have levels on both sides of the cursor. This may make local edits a bit unintuitive. An elegant solution is for the ends of the \raz{} to contain an ``empty'' element. Focusing on the far right element and inserting a new element to the left modifies the sequence more intuitively, and the empty element stays at the end.

\subsection{Enhancements}
One benefit of an extremely simple data structure is that it can be modified to suit a specific purpose. We annotate the \raz{} with size info in order to focus on a particular location. By using other types of annotations, the \raz{} can be used for other purposes. For example, by annotating each subtree with the highest priority of elements within, we have a priority tree, which can access the next item in log time, while still adding items in constant time. The paper on finger trees \cite{Hinze-Paterson:FingerTree} has additional suggestions of this kind.

Another modification might put the \raz{}'s cursor between two sequence elements for more intuitive editing. Each local editing function would require two cases, one for when there is a level present and one where there is an element. An invariant for which side the level appears on must be maintained to avoid degradation of the \raz{} structure.

It is possible for level data to be defined programmatically rather than randomly. This allows direct control over the balance of the unfocused \raz{}. For example, a perfect balance might be set for a read-only sequence. Or, one could use levels as priority controls. High levels mean faster access, so elements could be inserted along with a high priority where appropriate. Some additional design work needs to be done for this to work properly.

\section{Related Work and Alternative Approaches}
\label{related}

We review related work on representing purely-functional sequences
that undergo small (constant-sized) edits, supplementing the
discussions from earlier sections.
%
%
We also discuss hypothetical approaches based on (purely-functional)
search trees, pointing out their differences and short-comings for
representing sequences.

\paragraph{The ``Chunky Decomposition Scheme''.}
The tree structure of the \raz{} is inspired by the so-called ``chunky
decomposition scheme'' of sequences, from Pugh and Teiltelbaum's
1989~POPL~paper on purely-functional incremental
computing~\cite{PughTe89}.
Similar to skip lists~\cite{SkipLists89}, this decomposition scheme
hashes the sequence's elements to (uniquely) determine a
probabilistically-balanced tree structure.
The \raz{} enhances this structure with a focal point, local edits at
the focus, and a mechanism to overcome its inapplicability to
sequences of repeated (non-unique) elements.
In sum, the \raz{} admits efficient random access for
(purely-functional) \emph{sequence editing}, to which the '89 paper
alludes, but not does not address.

\paragraph{Finger trees.}
As introduced in \secref{intro}, a finger tree represents a sequence
and provides operations for a double-ended queue (aka, deque) that
push and pop elements to and from its two ends, respectively.
The 2-3 finger tree supports these operations in amortized constant time.
Structurally, it consists of nodes with three branches: a left branch
with 1--4 elements, a center for recursive nodes, and a right branch
with 1--4 elements.
Each center node consists of a \emph{complete} 2-3 tree.
%
%
This construction's shallow left and right (non-center) branches admit
efficient, amortized constant-time access to the deque's ends.
This construction also provides efficient (log-time) split
and append operations, for exposing elements in the center of the
tree, or merging two trees into a single 2-3 tree.
The split operation is comparable to the focus operation of the
\raz{}; the append operation is comparable to that of the \raz{}.

While similar in asymptotic costs, in settings that demand
\emph{canonical forms} and/or which employ \emph{hash consing}, the
2-3 finger tree and \raz{} are significantly different; this
can impact the asymptotics of comparing sequences for equality.
%
%
In the presence of hash-consing, structural identity coincides with
physical identity, allowing for $O(1)$-time equality checks of
arbitrarily long sequences.
As a result of their approach, 2-3 finger trees are history
dependent. This fact
makes them unsuitable for settings such as memoization-based
incremental computing~\cite{PughTe89,HammerKHF14,HammerDHLFHH15}.
%

\begin{figure}[h!]
  \includegraphics[width=\columnwidth]{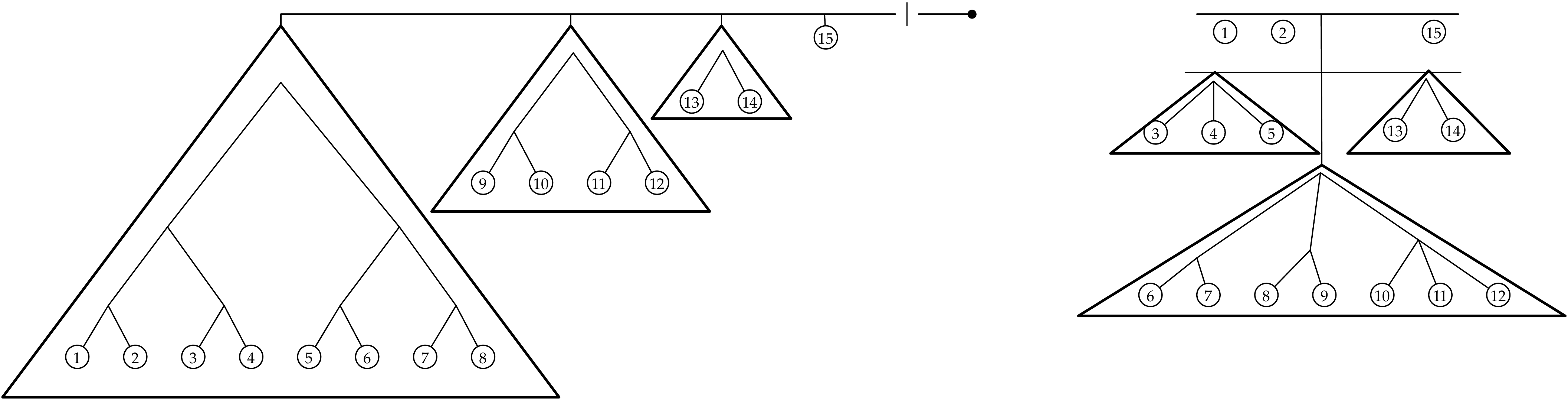}
  \caption{A \raz{} (left) and finger tree (right) representing the same sequence}
  \label{fig:fingertree1}
  \vspace{5mm}
  \includegraphics[width=\columnwidth]{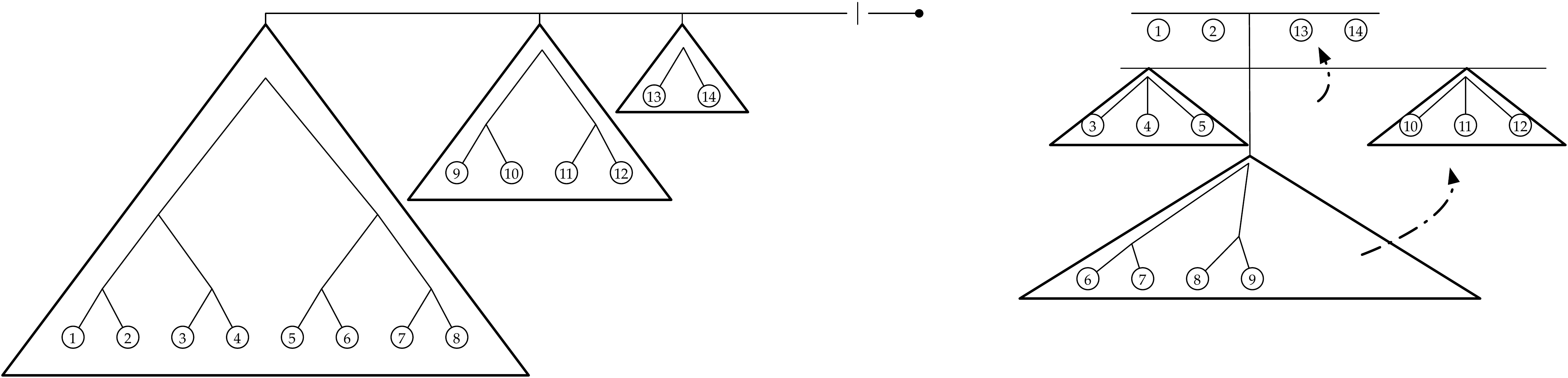}
  \caption{Removing an element from a \raz{} (left) and a finger tree (right), with structure maintenance on the finger tree}
\label{fig:fingertree2}
\end{figure}

\figref{fingertree1} depicts both a \raz{} and a finger tree
containing 15 elements, numbered sequentially. Elements are shown as
circles, internal trees have a triangle around them, with diagonal
lines denoting internal tree structure. Horizontal and vertical lines
elide a simple structure for access to elements: a list in the \raz{}
case and an set of data types for the finger tree. Both these data
structures provide access to the upper right element, labeled
``15''. We elide the current item and right side of the \raz{}, as it
is not important for this example.


One major difference between the finger tree and \raz{} is when they
need to adjust their structure to maintain invariants. Figure
\figref{fingertree2} shows the result of deleting element 15 in both
our example structures. They both expose this element, but the \raz{}
requires no maintenance at this point, while the finger tree does,
since there are no elements left in the top finger. This is done by
promoting a tree from the next deeper finger. In this case, the finger
tree must promote another tree from even deeper. These promotions are
indicated by the arrows in the figure.

\paragraph{RRB Vector.}
The \emph{{RRB}-Vector}~\cite{RRBVector} uses a balanced tree to
represent immutable vectors, focusing on practical issues such as
parallel performance and cache locality.
These performance considerations are outside the scope of our current
work, but are interesting for future work.

\paragraph{Balanced representations of Sets.}
Researchers have proposed many approaches for representing sets as
balanced search trees, many of which are amenable to
purely-functional representations (e.g., Treaps~\cite{Treaps89}, Splay
Trees~\cite{SplayTrees83}, AVL Trees~\cite{AVLTrees63}, and Red-Black
Trees~\cite{RedBlackTrees72}).
Additionally, skip lists~\cite{SkipLists89} provide a structure that
is tree-like, and which is closely related to the probabilistic
approach of the RAZ.
However, search trees (and skip lists) are \emph{not} designed to
represent sequences, but rather sets (or finite mappings).

Structures for sequences and sets are fundamentally different.
Structures for sequences admit operations that alter the presence,
absence and ordering of elements, and they permit elements to be
duplicated in the sequence (e.g., a list of $n$ repeated characters is
different from the singleton list of one such character).
By contrast, structures for sets (and finite maps) admit operations
that alter the presence or absence of elements in the structure, but
not \emph{the order} of the elements in the structure---rather, this
ordering is defined by the element's type, and is not represented by
the set.
Indeed, the set representation uses this (fixed) element ordering to
efficiently search for elements.
Moreover, set structures typically do not distinguish between the sets
with duplicated elements---e.g., \code{add(elm, set)} and
\code{add(elm, add(elm, set))} are the same set, whereas a sequence
structure would clearly distinguish these cases.

\paragraph{Encoding sequences with sets.}
In spite of these differences between sets and sequences, one can
\emph{encode} a sequence using a finite map, similar to how one can
represent a mutable array with an immutable mapping from natural
numbers to the array's content; however, like an array, editing this
sequence by element insertion and removal is generally problematic,
since each insertion or removal (naively) requires an $O(n)$-time
re-indexing of the mapping.
Overcoming this efficiency problem in turn requires employing
so-called \emph{order maintenance data structures}, which admit
(amortized) $O(1)$-time insertion, removal and comparison operations
for a (mutable) total order~\cite{DietzS87,BenderCDFZ02}.
Given such a structure, the elements of this order could be used
eschew the aforementioned re-indexing problem that arises from the
naive encoding of a sequence with a finite map.  Alas, existing order
maintenance data structures are \emph{not} purely-functional, so
additional accommodations are needed in settings that require
persistent data structures.
By contrast, the RAZ is simple, efficient and purely-functional.

\section{Conclusion}
We present the Random Access Zipper (\raz{}), a novel data structure for representing a sequence. We show its simplicity by providing most of the code, which contains a minimal number of cases and helper functions. We describe some of the design decisions that increase simplicity. We evaluate the \raz{}, demonstrating time bounds on par with far more complex data structures. Finally, we suggest multiple ways to enhance the \raz{} to suit additional use cases.

{\footnotesize
\bibliography{RAZ}

\begin{thebibliography}{10}

\bibitem{AVLTrees63}
M~AdelsonVelskii and Evgenii~Mikhailovich Landis.
\newblock An algorithm for the organization of information.
\newblock Technical report, DTIC Document, 1963.

\bibitem{Treaps89}
Cecilia~R. Aragon and Raimund Seidel.
\newblock Randomized search trees.
\newblock In {\em 30th Annual Symposium on Foundations of Computer Science,
  Research Triangle Park, North Carolina, USA, 30 October - 1 November 1989},
  pages 540--545, 1989.

\bibitem{RedBlackTrees72}
Rudolf Bayer.
\newblock Symmetric binary {B}-trees: Data structure and maintenance
  algorithms.
\newblock {\em Acta Inf.}, 1:290--306, 1972.

\bibitem{BenderCDFZ02}
Michael~A. Bender, Richard Cole, Erik~D. Demaine, Martin Farach{-}Colton, and
  Jack Zito.
\newblock Two simplified algorithms for maintaining order in a list.
\newblock In {\em Algorithms - {ESA} 2002, 10th Annual European Symposium,
  Rome, Italy, September 17-21, 2002, Proceedings}, pages 152--164, 2002.

\bibitem{DietzS87}
Paul~F. Dietz and Daniel~Dominic Sleator.
\newblock Two algorithms for maintaining order in a list.
\newblock In {\em Proceedings of the 19th Annual {ACM} Symposium on Theory of
  Computing, 1987, New York, New York, {USA}}, pages 365--372, 1987.

\bibitem{HammerDHLFHH15}
Matthew~A. Hammer, Joshua Dunfield, Kyle Headley, Nicholas Labich, Jeffrey~S.
  Foster, Michael~W. Hicks, and David~Van Horn.
\newblock Incremental computation with names.
\newblock In {\em Proceedings of the 2015 {ACM} {SIGPLAN} International
  Conference on Object-Oriented Programming, Systems, Languages, and
  Applications, {OOPSLA} 2015, part of {SPLASH} 2015, Pittsburgh, PA, USA,
  October 25-30, 2015}, pages 748--766, 2015.

\bibitem{HammerKHF14}
Matthew~A. Hammer, Yit~Phang Khoo, Michael Hicks, and Jeffrey~S. Foster.
\newblock Adapton: composable, demand-driven incremental computation.
\newblock In {\em {ACM} {SIGPLAN} Conference on Programming Language Design and
  Implementation, {PLDI} '14, Edinburgh, United Kingdom - June 09 - 11, 2014},
  page~18, 2014.

\bibitem{Hinze-Paterson:FingerTree}
Ralf Hinze and Ross Paterson.
\newblock Finger trees: A simple general-purpose data structure.
\newblock {\em Journal of Functional Programming}, 16(2):197--217, 2006.

\bibitem{Huet97}
G\'{e}rard Huet.
\newblock The zipper.
\newblock {\em Journal of Functional Programming}, 1997.

\bibitem{batteries-repo}
ocaml-batteries team.
\newblock Ocaml batteries included.
\newblock \url{https://github.com/ocaml-batteries-team/batteries-included}.
\newblock Accessed: 2016-07-12.

\bibitem{SkipLists89}
William Pugh.
\newblock Skip lists: {A} probabilistic alternative to balanced trees.
\newblock In {\em Algorithms and Data Structures, Workshop {WADS} '89, Ottawa,
  Canada, August 17-19, 1989, Proceedings}, pages 437--449, 1989.

\bibitem{PughTe89}
William Pugh and Tim Teitelbaum.
\newblock Incremental computation via function caching.
\newblock In {\em POPL}, 1989.

\bibitem{SplayTrees83}
Daniel~Dominic Sleator and Robert~Endre Tarjan.
\newblock Self-adjusting binary trees.
\newblock In {\em Proceedings of the 15th Annual {ACM} Symposium on Theory of
  Computing, 25-27 April, 1983, Boston, Massachusetts, {USA}}, pages 235--245,
  1983.

\bibitem{RRBVector}
Nicolas Stucki, Tiark Rompf, Vlad Ureche, and Phil Bagwell.
\newblock {RRB} vector: a practical general purpose immutable sequence.
\newblock In {\em {ICFP} 2015}, 2015.

\end{thebibliography}
\bibliographystyle{plain}
}

\end{document}